\documentclass[prb,preprint,showpacs,preprintnumbers,amsmath,amssymb]{revtex4}
\usepackage{bm}
\usepackage{epsfig}

\begin{document}
\title{Superconductivity from four Fermion complexes}

\author{P. Kleinert}
%\email[kl@pdi-berlin.de]{Your e-mail address}
\affiliation{Paul-Drude-Intitut f\"ur Festk\"orperelektronik,
Hausvogteiplatz 5-7, 10117 Berlin, Germany}
\date{\today}% It is always \today, today,
             %  but any date may be explicitly specified

\begin{abstract}
Superconductivity is studied for a fermionic system with
attractive four-body interaction. Applying a Green function
approach, the gap equation is derived. From the solution, the
transition temperature is calculated. Under the condition that the
respective coupling constants are comparable, the transition
temperature of four-fermion complexes is considerably larger than
the corresponding BCS value.
\end{abstract}


\maketitle

\section{Introduction}
About fifty years ago, the development of the BCS theory
\cite{bcs} and its overwhelming success in physics and application
was a breakthrough in modern theoretical physics. This theory
provides basic insight into superconductivity from a
quantum-mechanical point of view. According to the BCS mechanism,
the superconducting phase is due to the condensation of bosonic
quasi-particles, called Cooper pairs, which are created from
Fermions that experience an effective weak attractive interaction.
An essential ingredient of the BCS theory was the observation
\cite{Bardeen} that there could be a phonon-mediated attraction
between electrons in some metals at low temperatures. This
effective force acts at longer distances than the short-range
screened Coulomb interaction. In the simplified version of the
model \cite{bcs}, the attractive interaction is constant in a
narrow energy region (of the order of the Debye energy
$\hbar\omega_D$) around the Fermi energy and zero outside. As the
extent of a Cooper pair exponentially increases with decreasing
coupling, a large number of Cooper pair wave functions usually
overlap with each other. In the extreme limit of vanishing
attractive coupling $g_2$, the non-perturbative BCS theory becomes
exact.

The attractive interaction between fermions may originate from
quite different sources. Since the effective attraction could be
due not only to phonons but also to other bosonic excitations, the
BCS approach has been successfully applied to many quite different
research fields. Unfortunately, the basic physics of
high-temperature superconductivity \cite{Bednorz} does not fit
into any weak-coupling schema and therefore also not into the BCS
theory. The measured transition temperature $T_c$ in cuprates,
which can reach 160~K, cannot be derived from realistic BCS
calculations.

Regardless of this disadvantage, it is not exaggerated to state
that the BCS theory has deepened our understanding of many-body
effects in the quantum world. The non-perturbative character of
the approach, as revealed by the highly nonlinear dependence of
the transition temperature or the gap energy on the coupling
constant $g_2$, expresses the fact that the superconducting phase
is to be understood as a complex many-body phenomenon, to which an
infinite number of particular particle correlations contribute. In
addition, the resulting Cooper pairs, which carry the
superconducting current, do not exist as spatially separated
entities but are collected into large drops, in which novel
collective quantum effects could exist in principle. However, the
BCS approach focuses on pair correlations meaning that the
behavior of more than two highly correlated quantum objects is
described by two-body interactions between all possible pairs. The
question arises whether this first approximation is exhaustive for
the description of many-body quantum effects. In fact, many-body
interactions have been studied in different research areas like
nuclear, atomic, and condensed matter physics. The main reason
usually is the treatment of cluster effects created by an ensemble
of strongly correlated particles (especially self-consistent
cluster approximations \cite{Elliott}). In the field of nuclear
physics there is growing evidence that three-body forces exist
among the nucleons inside atomic nuclei. In condensed-matter
physics, many-body effects of trions were studied~\cite{Comb} and
three-body forces were shown to dominate the Cauchy discrepancies
in the second and third order elastic constants of
Copper.~\cite{Cousins} Three-body contributions to the interatomic
potential also modify many properties of the liquid and solid
phases of $^{4}$He atoms.~\cite{Ujevic} Even a spectacular result
have been reported in the field of many-body interaction
\cite{Date} namely that a $D$-dimensional Hamiltonian with two-
and three-body interactions has a unique string-like ground state,
when the strength of the three-body coupling exceeds a critical
value. In a recent study of the Bose-Einstein condensate-BCS
crossover in fermionic systems \cite{Mora}, the Schr\"odinger
equation for four attractively interacting fermions was solved by
applying Bethe ansatz techniques. The solution reveals a robust
four-fermion cluster that is not broken by collision and does not
couple to additional fermionic states.

The mentioned papers initiated a study of the rich and interesting
quantum physics that refers to many-body interaction. The
objective of the present paper is to contribute to this
fascinating field by treating a BCS-like instability in an
electron gas with four fermion coupling.

\section{Basic approach}
The BCS theory is based on the pair approximation with an
effective particle-particle coupling that is attractive and phonon
mediated. Similarly, it is conceivable that also a cluster of
three fermions is glued together by an attracting three-body
coupling. However, due to its fermionic character, these
aggregates are not expected to form a condensate. One needs two or
four fermion complexes for their transmutation into boson-like
objects, which could give rise to a phase transition to a
superconductor. Here, we focus on an instability, which is due to
the conglomerate of four fermions described by an effective
potential $v$. In order to concentrate on possible instabilities
in the system, both pair- and higher-order particle interactions
are not taken into account. Therefore, we start from the model
Hamiltonian
\begin{eqnarray}
&&H=\sum\limits_{s}\int d{\bm{r}} \psi_{s}^{\dag}({\bm{r}},t)
\left[-\frac{\hbar^2}{2m}\triangle_{\bm{r}}+V({\bm r}) \right]
\psi_{s}({\bm{r}},t)\\
&&+\sum\limits_{\{s_{i}\}}\int \{d{\bm{r}_{i}}\}
\psi_{s_1}^{\dag}({\bm{r}_1},t) \psi_{s_2}^{\dag}({\bm{r}_2},t)
\psi_{s_3}^{\dag}({\bm{r}_3},t)
\psi_{s_4}^{\dag}({\bm{r}_4},t)\nonumber\\
&&\times v({\bm{r}_1},{\bm{r}_2},{\bm{r}_3},{\bm{r}_4})
\psi_{s_4}({\bm{r}_4},t) \psi_{s_3}({\bm{r}_3},t)
\psi_{s_2}({\bm{r}_2},t) \psi_{s_1}({\bm{r}_1},t),\nonumber
\end{eqnarray}
where $V({\bm{r}})$ and $m$ denote the crystal potential and the
effective mass, respectively. Fermions with spin $s$ are created
[annihilated] by the operators $\psi_{s}^{\dag}({\bm{r}},t)$
[$\psi_{s}({\bm{r}},t)$]. Furthermore, the abbreviation
$\{d{\bm{r}_{i}}\}=d{\bm{r}_{1}}d{\bm{r}_{2}}d{\bm{r}_{3}}d{\bm{r}_{3}}$
is used. The one-particle Green function satisfies the equation of
motion \cite{Baym}
\begin{eqnarray}
&&\biggl\{i\hbar\frac{\partial}{\partial t_1}+
\frac{\hbar^2}{2m}\triangle_{\bm{r}_1}-V({\bm{r}_1})-U(1)
\biggl\}G(1,1^{\prime};U)=\delta (1-1^{\prime})\label{Baym}\\
&&-i\hbar^3\int d\overline{2} d\overline{3} d\overline{4}
V(1,\overline{2},\overline{3},\overline{4})G(1,\overline{2},\overline{3},\overline{4};
1^{\prime},\overline{2}^{+},\overline{3}^{+},\overline{4}^{+};U),\nonumber
\end{eqnarray}
with $U(1)\equiv U_s({\bm{r}},t)$ being an arbitrary auxiliary
function. In this equation, the usual notation is used, e.g.
$\overline{2}^{+}=\{{\bm{r}_{\overline{2}}},t_{\overline{2}}+i\varepsilon,s_{\overline{2}}
\}$ with $\varepsilon\rightarrow 0$. As Eq.~(\ref{Baym}) is not
closed, it must be supplemented by an equation for the
four-particle Green function and so on. This hierarchy is
alternatively deduced from the generating functional
$G[\lambda,\eta ]$
\begin{equation}
G(1,\dots,n;1^{\prime},\dots,n^{\prime};U)=\left.
\frac{\delta}{\delta\eta(n^{\prime})}\dots
\frac{\delta}{\delta\eta(1^{\prime})}\frac{\delta}{\delta\lambda(1)}
\dots \frac{\delta}{\delta\lambda(n)}G[\lambda,\eta]
\right|_{\lambda=\eta=0},
\end{equation}
by taking functional derivatives with respect to the Grassmannian
source fields $\lambda$ and $\eta$ before they are set equal to
zero. Expressing Eq.(\ref{Baym}) by a functional differential
equation
\begin{eqnarray}
&&\biggl\{i\hbar\frac{\partial}{\partial t_1}+
\frac{\hbar^2}{2m}\triangle_{\bm{r}_1}-V({\bm{r}_1})-U(1) \biggl\}
\frac{\delta}{\delta\lambda(1)}
G[\lambda,\eta]=\eta(1)G[\lambda,\eta]\label{Gen}\\
&&-i\hbar^3\int d\overline{2}d\overline{3}d\overline{4}
V(1,\overline{2},\overline{3},\overline{4})
\frac{\delta}{\delta\eta(\overline{4}^{+})}\frac{\delta}{\delta\eta(\overline{3}^{+})}
\frac{\delta}{\delta\eta(\overline{2}^{+})}\frac{\delta}{\delta\lambda(\overline{2})}
\frac{\delta}{\delta\lambda(\overline{3})}\frac{\delta}{\delta\lambda(\overline{4})}
\frac{\delta}{\delta\lambda(1)}G[\lambda,\eta],\nonumber
\end{eqnarray}
the coupled set of equations for the Green functions is easily
obtained by calculating additional functional derivatives of
desired order. At this stage, it is recommended to introduce the
generating functional $G_c[\lambda,\eta]$ for the connected parts
of many-particle Green functions via the equation
\begin{equation}
G[\lambda,\eta]=\exp\left\{ G_c[\lambda,\eta]\right\}.\label{e5}
\end{equation}
For instance, for the one- and two-particle functions, the
relationship
\begin{equation}
G(1,1^{\prime};U)=G_c(1,1^{\prime};U)\label{e6}
\end{equation}
\begin{equation}
G(1,2;1^{\prime},2^{\prime};U)=G_c(1,2;1^{\prime},2^{\prime};U)
+G(1,1^{\prime};U)G(2,2^{\prime};U)-G(1,2^{\prime};U)G(2,1^{\prime};U),\label{e7}
\end{equation}
is derived, from which it is concluded that the correlated
two-particle Green function does not enclose the Hartree-Fock
contributions. It is a general property that the correlated Green
functions vanish, when the interaction is absent so that the
hierarchy is suitably truncated by disregarding higher-level
many-particle contributions described by correlated Green
functions.

Inserting the correlated four-particle Green function into
Eq.~(\ref{Baym}), we obtain an equation, in which numerous terms
appear that represent various kinds of four-fermion scattering. It
is in line with our approach, which looks for a BCS-like
instability for four fermion complexes, to neglect all two- and
three-particle connected Green functions. The remaining
conglomerate of one-particle Green functions is collected into an
effective function $G_0(1,1^{\prime};U)$ [in the pair
approximation, which is the basis of the BCS theory, $G_0$ is
given by the Hartree-Fock Green function] so that we obtain
\begin{eqnarray}
G(1,1^{\prime};U)=&& G_0(1,1^{\prime};U)-i\hbar^3\int
d\overline{1}d\overline{2}d\overline{3}d\overline{4}
G_0(1,\overline{1};U)V(\overline{1},\overline{2},\overline{3},\overline{4})\nonumber\\
&&\times G_c(\overline{1},\overline{2},\overline{3},\overline{4};
1^{\prime},\overline{2}^{+},\overline{3}^{+},\overline{4}^{+};U).\label{eq1}
\end{eqnarray}
This result is complemented by the equation of motion for the
eight-point Green function, which is derived from Eqs. (\ref{Gen})
and (\ref{e5}). The handling of this rather complicated integral
equation is guided by the well established pair approximation
\cite{Puff, Kleinert} that leads to the BCS theory. In accordance
with the BCS reasoning, we focus on the homogeneous Cooper
channel, in which the connected four-particle function is
self-consistently determined from
\begin{eqnarray}
&&G_c(1,2,3,4;1^{\prime},2^{\prime},3^{\prime},4^{\prime};U)=
i\hbar^3 \int d\overline{1}d\overline{2}d\overline{3}d\overline{4}
G_0(1,\overline{1};U)V(\overline{1},\overline{2},\overline{3},\overline{4})\label{eq2}\\
&&\times
G(2,\overline{2}^{+};U)G(3,\overline{3}^{+};U)G(4,\overline{4}^{+};U)
G_c(\overline{1},\overline{2},\overline{3},\overline{4};1^{\prime},2^{\prime},3^{\prime},4^{\prime};U).
\nonumber
\end{eqnarray}
%%%%%%%%%%%%%%%%%%%%%%%%%%%%%%%
\begin{figure}[htbp]
%%\vspace*{5cm}
\epsfig{file=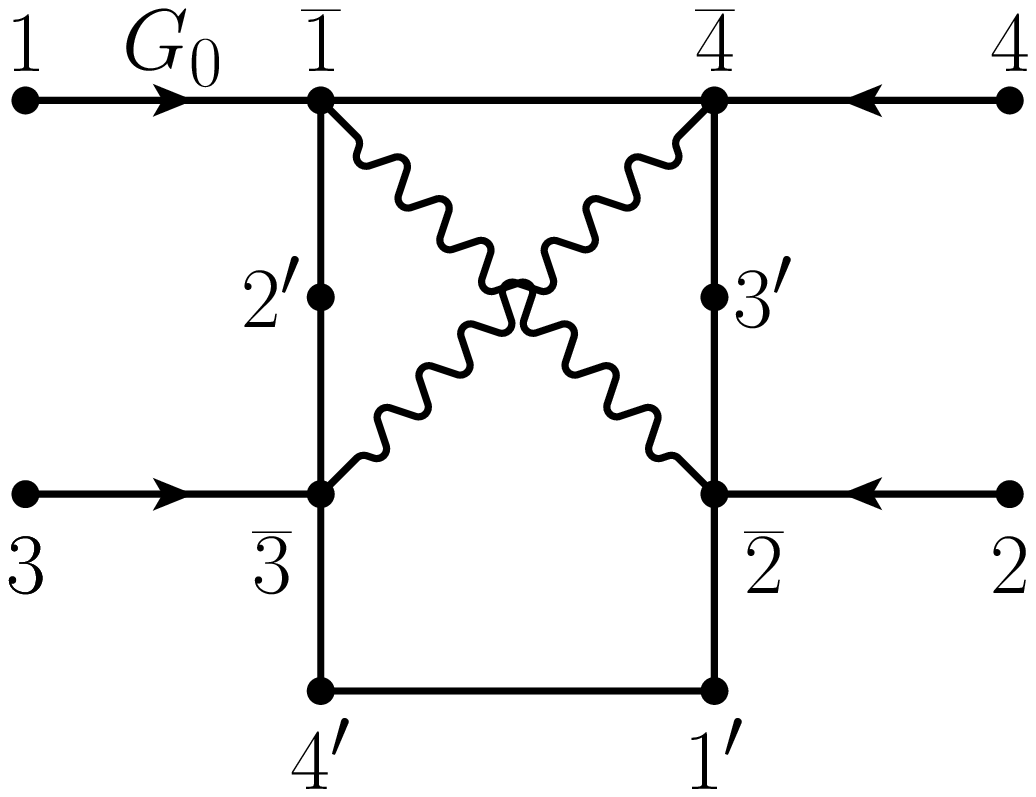,width=7.0cm,angle=0}%
\caption{Diagrammatic representation of the Cooper channel. The
wavy lines represent the four-body interaction.} \label{fig1}
\end{figure}
% %%%%%%%%%%%%%%%%%%%%%%%%%%%%%
A diagrammatic representation of the scattering contribution is
shown in Fig.~1. The basic set of Eqs.~(\ref{eq1}) and (\ref{eq2})
extends the BCS theory \cite{Puff, Kleinert} to fermions coupled
via a four-body potential. The solution of these equations is
facilitated by a Fourier transformation in the time domain.
Introducing the Matsubara frequencies
($z_{\nu}=i\pi\nu/(\hbar\beta)+\mu/\hbar$, $\nu=\pm 1,\pm3,\dots$
and $\omega_n=i\pi n/(\hbar\beta)+4\mu/\hbar$, $n=0,\pm 2,\dots$
with $\mu$ being the chemical potential and $\beta=1/k_BT$), the
Fourier transformation is defined by
\begin{eqnarray}
&&G_c(1,2,3,4;1^{\prime},2^{\prime},3^{\prime},4^{\prime};U)=
\left(\frac{i}{\hbar\beta} \right)^7 \sum\limits_{\{\nu_i\}}
\sum\limits_{\{\nu_i^{\prime}\}} \sum\limits_{n} G_c(
\{x_{i}\};\{x_{i}^{\prime}\}|
\{z_{\nu_{i}}\};\{z_{\nu_{i}^{\prime}} \}|\omega_n)\nonumber\\
&&\times \exp\left[-iz_{\nu_1}t_1-iz_{\nu_2}t_2-iz_{\nu_3}t_3
-i(\omega_n-z_{\nu_1}-z_{\nu_2}-z_{\nu_3})t_4 \right]\nonumber\\
&&\times
\exp\left[iz_{\nu_1^{\prime}}t_1^{\prime}+iz_{\nu_2^{\prime}}t_2^{\prime}+iz_{\nu_3^{\prime}}t_3^{\prime}
+i(\omega_n-z_{\nu_1^{\prime}}-z_{\nu_2^{\prime}}-z_{\nu_3^{\prime}})t_4^{\prime}
\right],\label{Fou}
\end{eqnarray}
where the spin and spatial coordinates are expressed by
$\{x_{i}\}=x_1,x_2,x_3,x_4$ with $x_j={\bm{r}_j},s_j$ and
$\{z_{\nu_{i}}\}=z_{\nu_1},z_{\nu_2},z_{\nu_3}$. When treating the
Fourier transformed basic Eqs. (\ref{eq1}) and (\ref{eq2}), the
following functions occur
\begin{equation}
G_c(\{x_{i}\};\{x_{i}^{\prime}\}|\{z_{\nu_{i}^{\prime}}
\}|\omega_n)= \left(
\frac{i}{\hbar\beta}\right)^3\sum\limits_{\{\nu_{i}\}}
G_c(\{x_{i}\};\{x_{i}^{\prime}\}|\{z_{\nu_{i}}
\};\{z_{\nu_{i}^{\prime}} \}|\omega_n),
\end{equation}
\begin{equation}
G_c(\{x_{i}\};\{x_{i}^{\prime}\}|\omega_n)= \left(
\frac{i}{\hbar\beta}\right)^6\sum\limits_{\{\nu_{i}\},\{\nu_{i}^{\prime}\}}
G_c(\{x_{i}\};\{x_{i}^{\prime}\}|\{z_{\nu_{i}}
\};\{z_{\nu_{i}^{\prime}} \}|\omega_n),
\end{equation}
\begin{equation}
\Delta(\{x_{i}\};\{x_{i}^{\prime}\}|\omega_n)=v(\{{\bm{r}_i} \})
v(\{{\bm{r}_i^{\prime}} \})
G_c(\{x_{i}\};\{x_{i}^{\prime}\}|\omega_n).
\end{equation}
The integral equation for the gap function $\Delta$ in the
representation with Matsubara frequencies is straightforwardly
obtained from Eq.~(\ref{eq2})
\begin{eqnarray}
&&\Delta(\{x_{i}\};\{x_{i}^{\prime}\}|\omega_n)= -i\hbar^3
v({\bm{r}_1},\dots,{\bm{r}_4})\int\{dx_{\overline{i}}
\}\left(\frac{i}{\hbar\beta} \right)^{3}\sum\limits_{\{\nu_{i} \}}
G_0(x_1,x_{\overline{1}},z_{\nu_1})G(x_2,x_{\overline{2}},z_{\nu_2})
\nonumber\\
&&\times G(x_3,x_{\overline{3}},z_{\nu_3})
G(x_4,x_{\overline{4}},\omega_n-z_{\nu_1}-z_{\nu_2}-z_{\nu_3})
\Delta(\{x_{\overline{i}}\};\{x_{i}^{\prime}\}|\omega_n).\label{e14}
\end{eqnarray}
To derive the corresponding Fourier transformed version of
Eq.~(\ref{eq1}), we account for the symmetry property
\begin{equation}
G_c(\{x_{i}\};\{x_{i}^{\prime}\}|\{z_{\nu_{i}}
\};\{z_{\nu_{i}^{\prime}} \}|\omega_n)=-
G_c^{*}(\{x_{i}^{\prime}\};\{x_{i}\}|\{z_{\nu_{i}^{\prime}}^{*}
\};\{z_{\nu_{i}}^{*} \}|\omega_n^{*}),
\end{equation}
which leads to an alternative formulation of Eq.~(\ref{e14}).
Inserting this gap function into Eq.~(\ref{eq1}) and performing a
spatial Fourier transformation by using a definition similar to
Eq.~(\ref{Fou}), we obtain
\begin{eqnarray}
&&G({\bm{k}},z_{\nu})=G_0({\bm{k}},z_{\nu})+\hbar^6G_0^2({\bm{k}},z_{\nu})
\left(\frac{i}{\hbar\beta}\right)\sum\limits_{{\bm{q}},n}\Delta({\bm{q}},\omega_{n})\label{a1}\\
&&\times\left(\frac{i}{\hbar\beta}
\right)^2\sum\limits_{{\bm{k}_1},\nu_1}
\sum\limits_{{\bm{k}_2},\nu_2}G({\bm{k}_1},z_{\nu_1})G({\bm{k}_2},z_{\nu_2})
G({\bm{q}}-{\bm{k}}-{\bm{k}_1}-{\bm{k}_2};\omega_n-z_{\nu}-z_{\nu_1}-z_{\nu_2}),
\nonumber
\end{eqnarray}
where the new gap function is given by
\begin{equation}
\Delta({\bm{q}},\omega_{n})=\sum\limits_{\{s_{i}
\}}\sum\limits_{\{{\bm{k}}_{i}
\}}\sum\limits_{\{{\bm{k}}_{i}^{\prime} \}} \Delta_{\{s_{i}
\},\{s_{i} \}}(\{{\bm{k}_{i}} \},\{{\bm{k}_{i}^{\prime}}
\}|{\bm{q}},\omega_{n}).
\end{equation}
In the derivation, it was considered
$G_{ss^{\prime}}=\delta_{ss^{\prime}}G$, which dictates the spin
dependence of the gap function. The basic Eq.~(\ref{a1}) for the
one-particle Green function provides an obvious extension of the
BCS singlet state description.~\cite{Puff,Kleinert} Similarly, we
obtain for the gap equation the following Fourier-transformed
version
\begin{eqnarray}
&&\Delta(\{{\bm{k}_i} \};\{{\bm{k}_i}^{\prime}
\}|{\bm{q}},\omega_n)=-i\hbar^3\sum\limits_{\{{\bm{k}_{\overline{i}}}
\}}
v({\bm{k}_1}-{\bm{k}_{\overline{1}}},{\bm{k}_1}-{\bm{k}_{\overline{1}}}
+{\bm{k}_2}-{\bm{k}_{\overline{2}}}\nonumber\\
&&,{\bm{k}_1}-{\bm{k}_{\overline{1}}}
+{\bm{k}_2}-{\bm{k}_{\overline{2}}}+{\bm{k}_3}-{\bm{k}_{\overline{3}}})
\left(\frac{i}{\hbar\beta} \right)^3\sum\limits_{\{ \nu_i\}}
G_0({\bm{k}_{\overline{1}}},z_{\nu_1})
G({\bm{k}_{\overline{2}}},z_{\nu_2})
G({\bm{k}_{\overline{3}}},z_{\nu_3})\nonumber\\
&&\times
G({\bm{q}}-{\bm{k}_{\overline{1}}}-{\bm{k}_{\overline{2}}}-{\bm{k}_{\overline{3}}}
,\omega_n-z_{\nu_1}-z_{\nu_2}-z_{\nu_3})\Delta(\{{\bm{k}_{\overline{i}}}
\};\{{\bm{k}_i}^{\prime} \}|{\bm{q}},\omega_n).\label{a2}
\end{eqnarray}
In the spirit of the BCS theory \cite{Puff,Kleinert},
Eqs.~(\ref{a1}) and (\ref{a2}) are solved in the limit
${\bm{q}}={\bm{0}}$ and $n=0$, which is an approximation
concerning the effective bosonic excitation composed of four
particles. In addition, it is assumed that the effective
attraction is given by a contact interaction, which is confined to
a narrow shell of width $2\hbar\omega_D$ around the Fermi energy.
Introducing the new abbreviations ${\bm{k}_2}+{\bm{k}_3}={\bm{q}}$
and $z_{\nu_2}+z_{\nu_3}=\omega_m$, we obtain the final set of
coupled equations
\begin{equation}
G({\bm{k}},z_{\nu})=G_0({\bm{k}},z_{\nu})+\Delta\hbar^6
G_0^{2}({\bm{k}},z_{\nu})\left(\frac{i}{\hbar\beta}
\right)\sum\limits_{{\bm{q}},m}u({\bm{q}},\omega_{m})
G^{*}({\bm{k}}+{\bm{q}},z_{\nu}+\omega_{m}),\label{b1}
\end{equation}
\begin{equation}
-i\hbar^3g_{4}\left(\frac{i}{\hbar\beta}
\right)\sum\limits_{{\bm{q}},m}u({\bm{q}},\omega_{m})
\left(\frac{i}{\hbar\beta} \right)\sum\limits_{{\bm{k}},\nu}
G_0({\bm{k}},z_{\nu})G^{*}({\bm{k}}+{\bm{q}},z_{\nu}+\omega_{m})=1,
\label{b2}
\end{equation}
\begin{equation}
u({\bm{q}},\omega_{m})=\left(\frac{i}{\hbar\beta}
\right)\sum\limits_{{\bm{k}_1},\nu_{1}}G({\bm{k}_1},z_{\nu_{1}})
G({\bm{q}}-{\bm{k}_1},\omega_m-z_{\nu_{1}}),\label{b3}
\end{equation}
which generalizes the BCS theory to a fermionic system with
four-body interaction. These equations allow a detailed
description of the superconducting phase of the considered cluster
model. Formally, Eqs.~(\ref{b1}) and (\ref{b2}) reduce to the
basic BCS equations, when the function $u({\bm{q}},\omega_m)$ is
replaced by $-i\hbar\beta\delta_{m,0}\delta_{{\bm{q}},{\bm{0}}}$.

\section{Transition temperature}
The internal structure of the four-particle cluster is mainly
accounted for by the function $u({\bm{q}},\omega_n)$, which has a
bosonic character. To study its influence in more detail, let us
treat the transition temperature $T_c$, at which the gap energy
$\Delta$ vanishes. For the calculation of $T_c$, the
self-consistent one-particle Green function $G$ is given by $G_0$,
for which, in analogy to the BCS approach, the simple expression
\begin{equation}
G_0({\bm{k}},z_{\nu})=\frac{1}{\hbar
z_{\nu}-\varepsilon_{{\bm{k}}}},
\end{equation}
is adopted. $\varepsilon_{\bm{k}}$ denotes the kinetic energy
$\hbar^2{\bm{k}^2}/2m$. For the function $u({\bm{q}},\omega_n)$,
we immediately obtain
\begin{equation}
u({\bm{q}},\omega_{l})=\frac{i}{\hbar}\sum\limits_{\bm{k}}
\frac{n_F(\xi_{{\bm{k}}})+n_F(\xi_{{\bm{q}}-{\bm{k}}})-1}{i\omega_l
-\xi_{{\bm{k}}}-\xi_{{\bm{q}}-{\bm{k}}}},
\end{equation}
with the Fermi distribution function $n_F$, $i\omega_{l}=2\pi i
l/\beta_c$, $\beta_c=1/k_BT_c$, and
$\xi_{{\bm{k}}}=\varepsilon_{\bm{k}}-\mu$. Inserting this result
into Eq.~(\ref{b2}), we obtain the gap equation
\begin{eqnarray}
&&g_4\sum\limits_{\bm{k}_1,\bm{k}_2}\sum\limits_{\bm{q}}\left[
n_F(\xi_{\bm{k}_1+\bm{q}})+n_F(\xi_{\bm{k}_1})-1 \right] \left[
n_F(\xi_{\bm{k}_2-\bm{q}})+n_F(\xi_{\bm{k}_2})-1
\right]\nonumber\\
&&\times\frac{n_B(\xi_{\bm{k}_1+\bm{q}}+\xi_{\bm{k}_1})+
n_B(\xi_{\bm{k}_2-\bm{q}}+\xi_{\bm{k}_2})+1}{\xi_{\bm{k}_1+\bm{q}}+
\xi_{\bm{k}_1}+\xi_{\bm{k}_2-\bm{q}}+\xi_{\bm{k}_2}}=1,\label{gap1}
\end{eqnarray}
in which the Bose distribution function $n_B$ and the four-body
coupling $g_4$ appear. Let us compare this result of the BCS-like
approach for four fermions, which are held together by a four-body
phonon-mediated attraction, with the gap equation of the BCS
theory
\begin{equation}
g_2\sum\limits_{\bm{k}}\frac{2n_F(\xi_{\bm{k}})-1}{2\xi_{\bm{k}}}=1,
\end{equation}
the solution of which is given by the well-known formula
\begin{equation}
T_c=1.13\frac{\hbar\omega_D}{k_B}\exp\left(-\frac{1}{g_2N_0}
\right),\label{TcBCS}
\end{equation}
which applies under the conditions that there is an attractive
pair interaction and that the inequalities $\mu\gg\hbar\omega_D$,
$\beta_c\hbar\omega_D/2\gg 1$ are satisfied. $N_0$ denotes the
density of states at the Fermi energy.
%%%%%%%%%%%%%%%%%%%%%%%%%%%%%%%
\begin{figure}[htbp]
%%\vspace*{5cm}
\epsfig{file=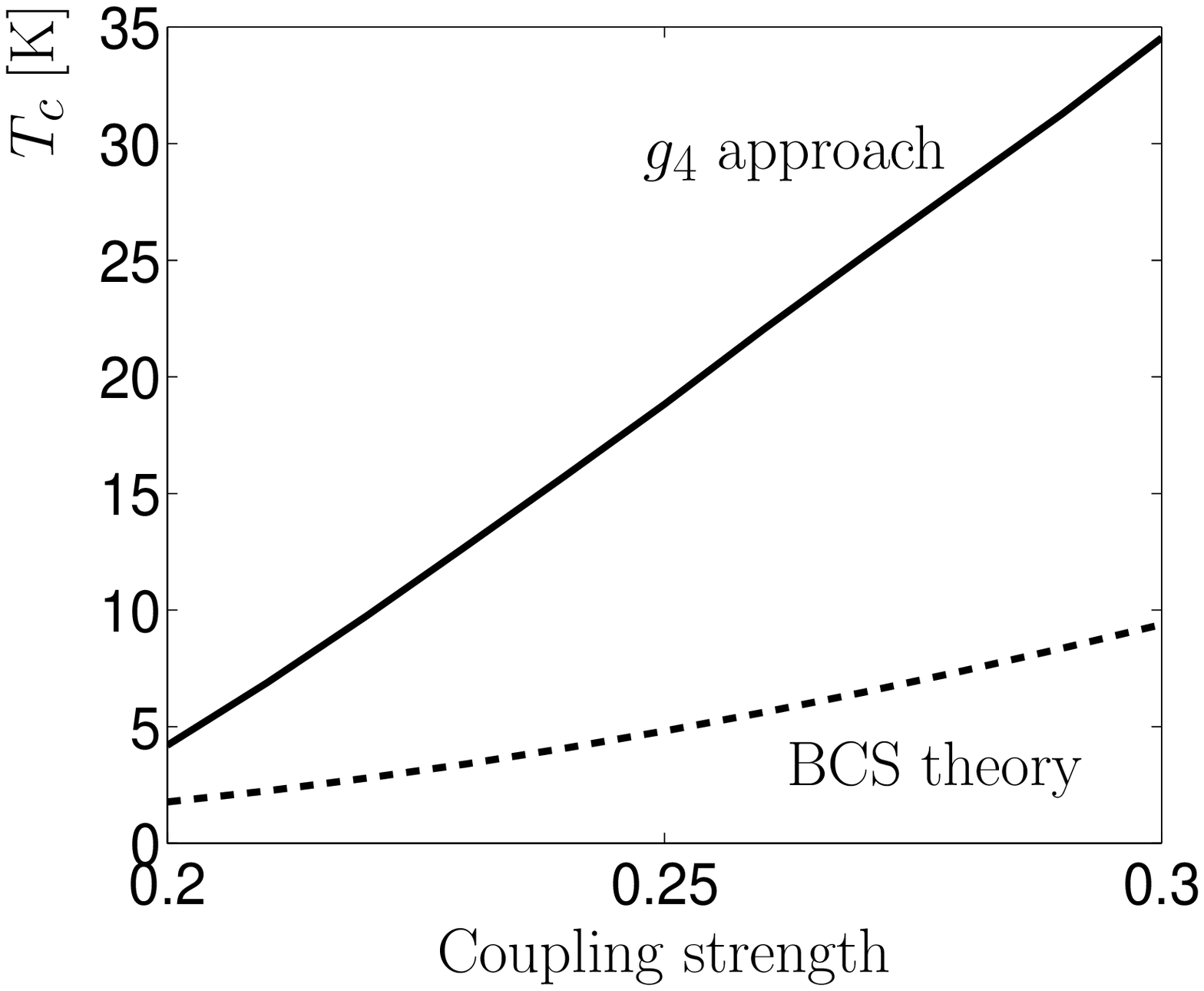,width=7.0cm,angle=0}%
\caption{Transition temperature for the $g_4$ approach (upper
curve) compared with the BCS result (lower curve) for
$\hbar\omega_D=0.02$~eV. The effective dimensionless coupling
strengths $g_2N_0$ and $g_4N_0^2\hbar\omega_D/4$ are assumed to be
equal.} \label{fig2}
\end{figure}
% %%%%%%%%%%%%%%%%%%%%%%%%%%%%%
To solve the gap Eq.~(\ref{gap1}) of the four-fermion cluster, it
is considered that the energies $\varepsilon_{{\bm{k}}_1}$ and
$\varepsilon_{{\bm{k}}_2}$ are confined to a narrow interval
around the Fermi surface so that the main contribution comes from
${\bm{q}}={\bm{0}}$. Adopting this approximation, we arrive at
\begin{equation}
\frac{g_4}{4}N_0^2\int\limits_{-\hbar\omega_D}^{\hbar\omega_D}
dE_1dE_2\tanh\left(\frac{\beta_c E_1}{2} \right)
\tanh\left(\frac{\beta_c E_2}{2} \right) \frac{\coth(\beta_c
E_1)+\coth(\beta_c E_2)}{E_1+E_2}=1.\label{gap2}
\end{equation}
The numerical solution of this equation for characteristic
coupling strengths of Cooperons is shown in Fig. 2 by the upper
curve and compared with the BCS result (lower curve). For the
comparison, the effective pair and four-body coupling constants
were set to be equal. In this case, the transition temperature of
four-fermion clusters is considerably higher than the BCS value.
Whether the predicted high-transition temperature is realistic
depends on possible values of the four-fermion coupling strength.
The study of this problem requires further work that goes beyond
the scope of our paper.

\section{Summary}
Guided by a Green function approach to BCS superconductivity, a
superconducting phase transition has been identified in a
fermionic system with a weakly attracting four-body interaction.
The basic building blocks of the superconducting phase are
quasi-particles that consist of four fermions hold together by an
effective attraction. Special results have been obtained for the
transition temperature $T_c$ by solving the gap equation. For
comparable coupling strengths, the transition temperature of the
four-fermion cluster model is much higher than the BCS values.
This kind of superconductivity should be observable
(preferentially by Andreev reflection) for strong coupling
strengths within the four-fermion conglomerate. Whether the
agglomerated attraction of four fermions can reach characteristic
BCS values is a question that will decide future studies.

%\bibliographystyle{prsty}
%\bibliography{abbrev,bcs}
%

%%%%%%%%%%%%%%%%%%%%%%%%%%%%%%%%%%%%%%%%%%%%%%%%%%%%%%%%%%%%%%%%%%%

\end{document}